\pdfoutput=1
\documentclass[aps,prl,twocolumn,showpacs,groupedaddress,\if{lengthcheck}\fi,
amsmath,amssymb,floatfix
]{revtex4}

\usepackage{epsf,
amsmath,graphicx,amssymb,subfigure}

\newcommand{\bs}{\mbox{\boldmath{$\sigma$}}}
\newcommand{\bS}{\mbox{\boldmath{$\Sigma$}}}
\begin{document}
\title{A Simple Model for the Deformation-Induced Relaxation \\
of Glassy Polymers}
\author{S. M. Fielding,$^1$  R. G. Larson,$^2$ M. E. Cates$^{3}$}
\affiliation{$^1$ Department of Physics, Durham University, Science Laboratories, South Road, Durham DH1 3LE, UK \\
$^2$ Department of Chemical Engineering, University of Michigan, Ann Arbor, MI 48109-2136, USA\\
$^3$ SUPA, School of Physics and Astronomy, The University of
  Edinburgh, JCMB Kings Buildings, Edinburgh, EH9 3JZ, UK}
\date{\today}
\begin{abstract}
{ 
Glassy polymers show ``strain hardening'': at constant extensional load, their flow first accelerates, then arrests. Recent experiments have found this to be accompanied by a striking and unexplained dip in the segmental relaxation time.  Here we explain such behavior by combining a minimal model of flow-induced liquefaction of a glass, with a description of the stress carried by strained polymers, creating a non-factorable interplay between aging and strain-induced rejuvenation. Under constant load, liquefaction of segmental motion permits strong flow that creates polymer-borne stress. This slows the deformation enough for the segmental modes to re-vitrify, causing strain hardening. 
}
\end{abstract}

\maketitle

Understanding the flow of polymeric materials is central to their manufacture and performance. After decades of progress, the flow properties of molten polymers are elegantly described by modern entanglement theories \cite{McLeishReview,LarsonBook}.  
In use, however, most polymeric materials are not molten, but rigid. This conversion is commonly achieved by cooling to below the glass transition temperature, $T_g$. In contrast to the molten case, satisfactory theories of polymer glass rheology remain elusive. 

Just below $T_g$, polymer glasses undergo slow plastic deformation if stress is applied \cite{Ediger,Struik}. 
Similar plasticity is shown also by molecular, metallic, and colloidal glasses \cite{Molecular,Metallic,Colloidal}.
Our understanding of flow in such non-polymeric glasses has improved greatly due to recent advances in microscopic \cite{BraderPRL2,BraderPNAS} and mesoscopic \cite{SGR,KS1,STZ} theory. Crucial to glass rheology is physical aging: a quiescent glass becomes more sluggish with time,  rejuvenating under flow. This is captured schematically in minimal `fluidity' models, with a time evolution equation for a single structural relaxation rate (the fluidity) \cite{Fluidity,Fluidity2}. In the so-called `simple aging' scenario, the structural relaxation time $\tau$ (or inverse fluidity) of the system at rest increases linearly with its age \cite{SGR,Aging,Struik}. A slow steady flow cuts off this growth at the inverse flow rate.

In polymeric glasses, new properties emerge from the interplay between polymeric and glassy degrees of freedom. Particularly striking is the evolution of the segmental relaxation time $\tau(t)$, controlling the rate of local rearrangements, when a load is applied. Recently, Lee et al \cite{Ediger,Ediger2} showed that $\tau(t)$ falls steadily during the early stages of elongational deformation, and then more sharply, reaching a small fraction $\sim 10^{-3.3}$  
of its initial level before dramatically rising again, as the local strain rate started to drop on entering the `strain hardening' regime. While elements of this scenario have been confirmed in coarse-grained and molecular simulations \cite{Simulations1,Simulations2,Simulations3,Simulations4}, no convincing theoretical picture has yet emerged. 

In \cite{Ediger}, the results for $\tau(t)$ were found inconsistent with the early theory of Eyring \cite{Eyring} and also with a more recent model \cite{KS1} (see also \cite{KS2,KS3,KS6}) involving broadly similar precepts. 
Indeed the Eyring-like assumption of a purely stress-dependent fluidity, as introduced for polymers in \cite{EGP2}, is fundamentally at odds with aging in glasses, whose fluidity is time-dependent at constant stress
\cite{Aging,SGR,Struik}. Previous work to incorporate aging and flow-rejuvenation into polymer glass theory has led to the Eindhoven Glassy Polymer model (EGP) \cite{EGP}, where viscosity is controlled by a state parameter $S$ that is age- and strain-dependent. However, in the EGP model treats aging and rejuvenation have factorable effects on $S$:  strain-induced rejuvenation causes cumulative losses of structure (reductions in $S$) which multiplicatively reduces all subsequent relaxation times. This
is not what theories of simple glasses predict
\cite{SGR,Fluidity,Fluidity2}. The EGP's precepts may thus be unsuited to the regime of strong fluidization, as addressed experimentally in \cite{Ediger} and in recent glass rheology theories \cite{BraderPRL2,BraderPNAS,SGR,Fluidity,Fluidity2}.

Despite several recent efforts \cite{KS1,KS2,KS6,hoy}, creating a comprehensive theory of rheological aging in polymer glasses remains a formidable task. 
Here we show that a minimal model, combining just two key elements of any such theory (nonfactorable aging/rejuvenation, and the strain dependence of polymer-borne stresses), semiquantitatively explains many of the results reported in \cite{Ediger}.  

Our model describes polymeric dumb-bells \cite{LarsonBook} suspended in a glassy `solvent', whose microscopic relaxation time obeys a fluidity-type equation showing simple aging and flow-rejuvenation.
Despite our nomenclature, we do not require any actual solvent to be present: the separation between polymer and `solvent' instead divides the slow degrees of freedom of large sections of chain from the shorter-scale and faster relaxing modes that control local segmental dynamics. 
Our model thus follows lines developed in \cite{EGP,EGP2} but crucially differs in its treatment of aging and rejuvenation.
For simplicity we treat the dumb-bells initially as purely elastic elements -- as is valid in the molten state, where the elasticity is of entropic origin \cite{LarsonBook}. However, we later return to discuss the true nature of the polymer stress in polymeric glasses which is not solely entropic in character \cite{KS6,Robbins}.

Our model first defines a deviatoric stress tensor $\bS = G^p(\bs^p-{\bf I}) + G^s(\bs^s-{\bf I})$ where $\bs^p$ and $\bs^s$ are dimensionless conformation tensors for polymer and `solvent', $G^{p,s}$ associated elastic moduli (see below), and ${\bf I}$ the unit tensor. 
We then adopt the following equations for the conformation tensors
and solvent relaxation time $\tau$:
\begin{eqnarray}
\dot\bs^p+{\bf v}.\nabla\bs^p &=& \bs^p.\nabla{\bf v} + (\nabla{\bf v})^T.\bs^p - \alpha(\bs^p - {\bf I})/\tau \label{one}
\\
\dot\bs^s+{\bf v}.\nabla\bs^s &=& \bs^s.\nabla{\bf v} + (\nabla{\bf v})^T.\bs^s - (\bs^s - {\bf I})/\tau \label{two}
\\
\dot\tau +{\bf v}.\nabla\tau &=& 1 - (\tau-\tau_0)\lambda 
\label{three}
\\
\lambda({\bf D}) &\equiv & \mu\sqrt{2\hbox{\rm Tr}({\bf D}.{\bf D})} \label{four}
\end{eqnarray}
Here ${\bf v}$ is the fluid velocity and ${\bf D} = (\nabla{\bf v}+(\nabla{\bf v})^T)/2$.

Eq.(\ref{one}) is an upper-convected Maxwell model \cite{LarsonBook}, describing the dynamics of elastic dumb-bells; these carry an elastic stress $G^p\bs^p$
and have a structural relaxation time $\tau^p=\tau/\alpha$, proportional to, but much larger than, that of the `solvent', $\tau$. In the simplest models of dense, molten, but unentangled polymers, $\alpha=N^{-2}$ with $N$ the polymerization index \cite{LarsonBook}, whereas in a lightly crosslinked elastomeric network  \cite{Ediger} one expects $\alpha = 0$.
Consistent with its glassy nature, the solvent itself is viewed as a viscoelastic fluid.  Bearing in mind that it represents shorter-scale polymeric degrees of freedom, we model this fluid using another upper-convected Maxwell model (\ref{two}). Because there are more local than chain-scale degrees of freedom, we expect $G^s > G^p$.

Finally, the solvent's structural relaxation time $\tau$ obeys a fluidity-type equation (\ref{three}), with the following two 
features. First, without flow, $\tau$ increases linearly in time at a (dimensionless) solidification rate $\dot\tau({\bf D = 0})$ which for simplicity we set to unity. This embodies the simple aging scenario that emerges from mesoscopic models \cite{SGR}, whereby 
local configurations evolve into ever deeper traps. 
Second, with flow present, $\tau$ would, in the absence of such aging, itself undergo deformation-induced relaxation towards $\tau_0$ which is a `fully rejuvenated' value. This relaxation occurs at a rate $\lambda$,  proportional to a scalar measure of flow rate (with $\mu$ another dimensionless coefficient \cite{LarsonBook,BraderPNAS}). 
In steady shear ($\lambda = \mu\dot\gamma$), $\tau$ then varies inversely with strain rate $\dot\gamma$ in accord with microscopic theory \cite{BraderPRL2}.
For uniaxial elongation at strain rate $\dot\varepsilon$, (\ref{four}) reduces to  $\lambda = \mu\sqrt{3}|\dot\varepsilon|$.
Note that in this simple fluidity model, the rejuvenation of $\tau$ is essentially strain-induced \cite{BraderPNAS} but, in contrast to the factorable model of \cite{EGP}, can be rapidly reversed by subsequent aging.

Our model is completed by the standard equations of mass and force balance for an incompressible fluid of negligible inertia:
$\nabla.{\bf v} = 0$, and
$\nabla.[\bS+ 2\eta{\bf D}] = 0$. (Here $\eta$ is a small additional Newtonian viscosity, included solely for numerical reasons \cite{Methods}.) We have solved our model numerically for uniaxial extension flows within a lubrication approximation appropriate to long cylindrical samples. Our numerical solutions address two cases \cite{Methods}. One is an effectively infinite cylinder that remains of spatially uniform, but time dependent, cross section. The second addresses a finite cylinder perturbed to trigger an inherent `necking' instability. The latter is commonplace in elongation of polymer glasses (and reported in \cite{Ediger}, in relatively mild form). We show next however that a semiquantitative account of the $\tau(t)$ response 
under elongational load is already predicted by applying our simple model to the infinite uniform cylinder.

In confronting the experimental data for $\tau(t)$ we first set $\alpha$ negligibly small, appropriate for a crosslinked material \cite{Ediger}.  The experimental protocol of \cite{Ediger} determines the applied tensile force $F$; the initial relaxation time ($t_w$ in our model) before applying the load; and the time $t_u$ at which unload later occurs. 
There remain four material parameters in the model: $G^p,G^s,\tau_0$ and $\mu$. As detailed in \cite{Methods}, three of these are strongly constrained by measurements that do not involve the dip in the $\tau(t)$ curve. Indeed, $G^p/F$ can be deduced from the asymptotic deformation in the strain-hardened regime just before unload; once $G^p$ is known (we find $G^p = 6$ MPa) $G^s$ and $\mu$ are in turn estimated from the step-change in $\tau$ during initial loading, and from the separately measured slope \cite{Ediger} of the `effective flow curve' $\dot\varepsilon(\tau)$. Hence the only unconstrained parameter 
in fitting the dip in $\tau(t)$
is $\tau_0$. 

\begin{figure}
\begin{center}
\includegraphics[width=86mm]{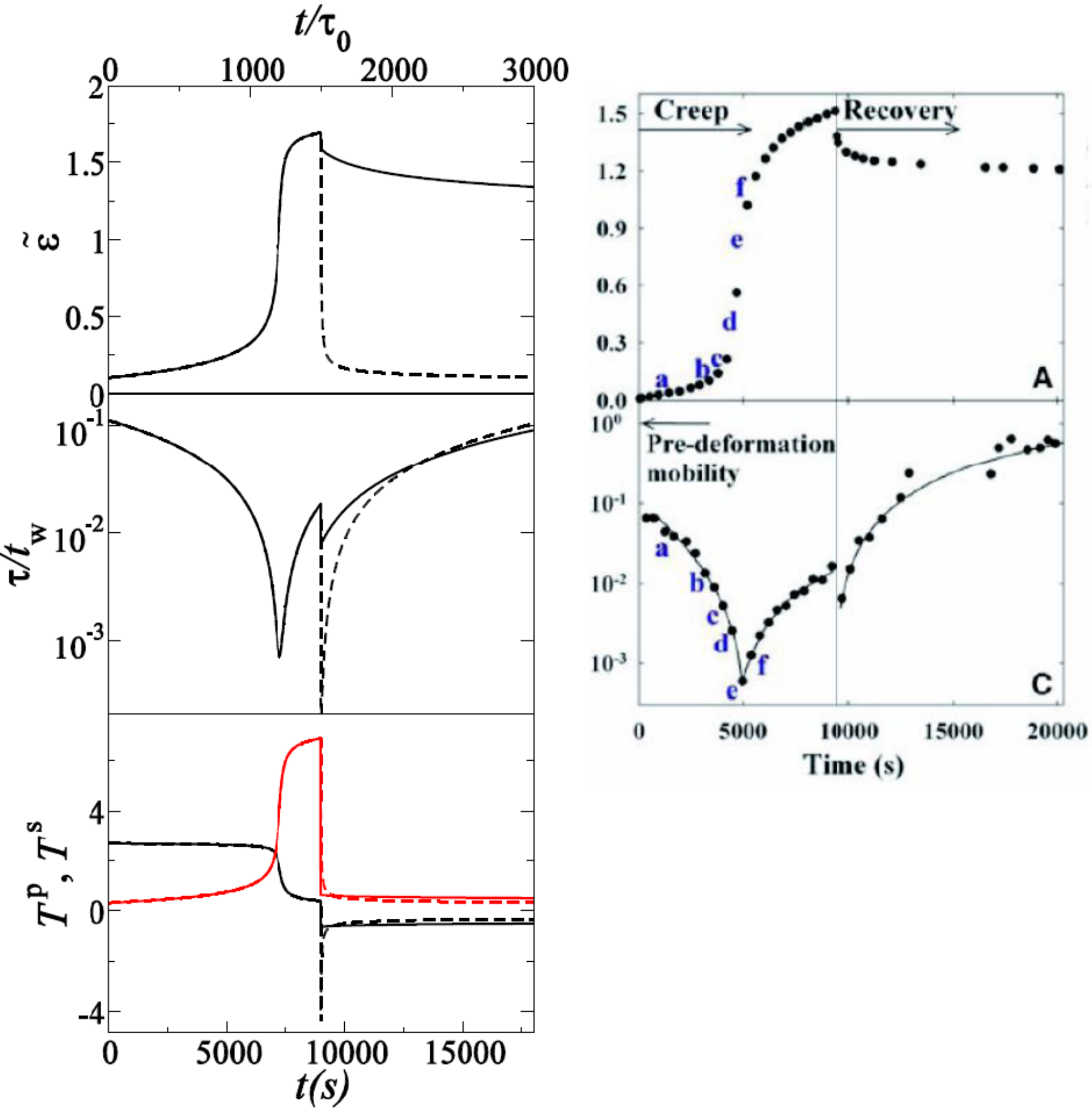}
\caption{ (Color online) Left Panels, solid curves: local strain
$\tilde{\varepsilon}=\exp\varepsilon-1$ \cite{Ediger}, reduced relaxation time $\tau(t)/t_w$ 
and tensile
stresses $T^{p,s}=G^{p,s}(\sigma_{zz}^{p,s}-\sigma_{xx}^{p,s})$ of the
polymer (p) and solvent (s) during loading of an infinite uniform
cylinder. Parameters $G^s/G^p =8.5$, $\mu=12.5$, $t_w/\tau_0=10^4$, $\tau_0=6$s; applied force / initial area $f=2.7G_{\rm p}$. 
(The curve for $T^p$, in red, initially lies below $T^s$ but crosses it during strain hardening.) 
The unload results for
the basic model ($\theta=1$) is shown dashed; the solid curve
after unload has $\theta = 0.1$. The horizontal axis is marked both in dimensionless model units (top) and real time (converted using $\tau_0$), bottom. Right Panels: Comparable experimental data for local strain and reduced relaxation time. (From ~\cite{Ediger}. Reprinted by permission of AAAS.)}
\label{fig:one}
\end{center}
\end{figure}

\begin{figure}
\begin{center}
\includegraphics[width=85mm]{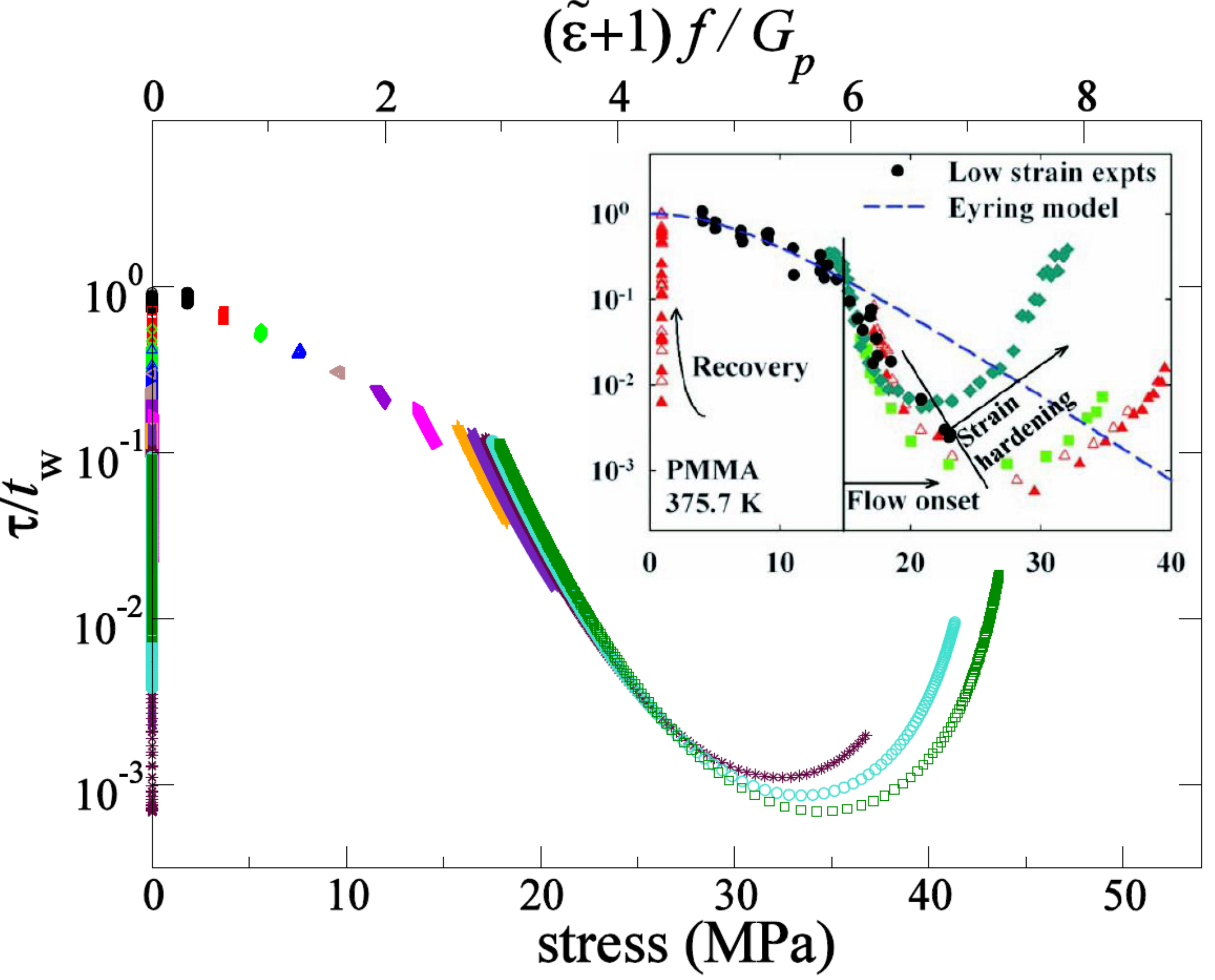}
\caption{(Color online) Reduced relaxation time $\tau/t_w$ against actual stress in
the infinite uniform cylinder. Parameter values $G^s/G^p =8.5$, $\mu=12.5$,
$t_w/\tau_0=10^4$, $\tau_0=6$s for a scaled applied force per initial area
$f/G_{\rm p}=0.3,0.6,0.9,1.2,1.5,1.8,2.1,2.4,2.5,2.6,2.65,2.7$ (increasing left to right). The unload time obeys $T_{\rm
unload}=1500\tau_0$. The horizontal axis is marked both in dimensionless model units (top) and laboratory stress (bottom, as used in the inset; converted factor $G^p = 6$ MPa). These curves show qualitative agreement with the experimental data (from ~\cite{Ediger}; reprinted by permission of AAAS) (inset).}
\label{fig:twoA}
\end{center}
\end{figure}

\begin{figure}
\begin{center}
\includegraphics[width=85mm]{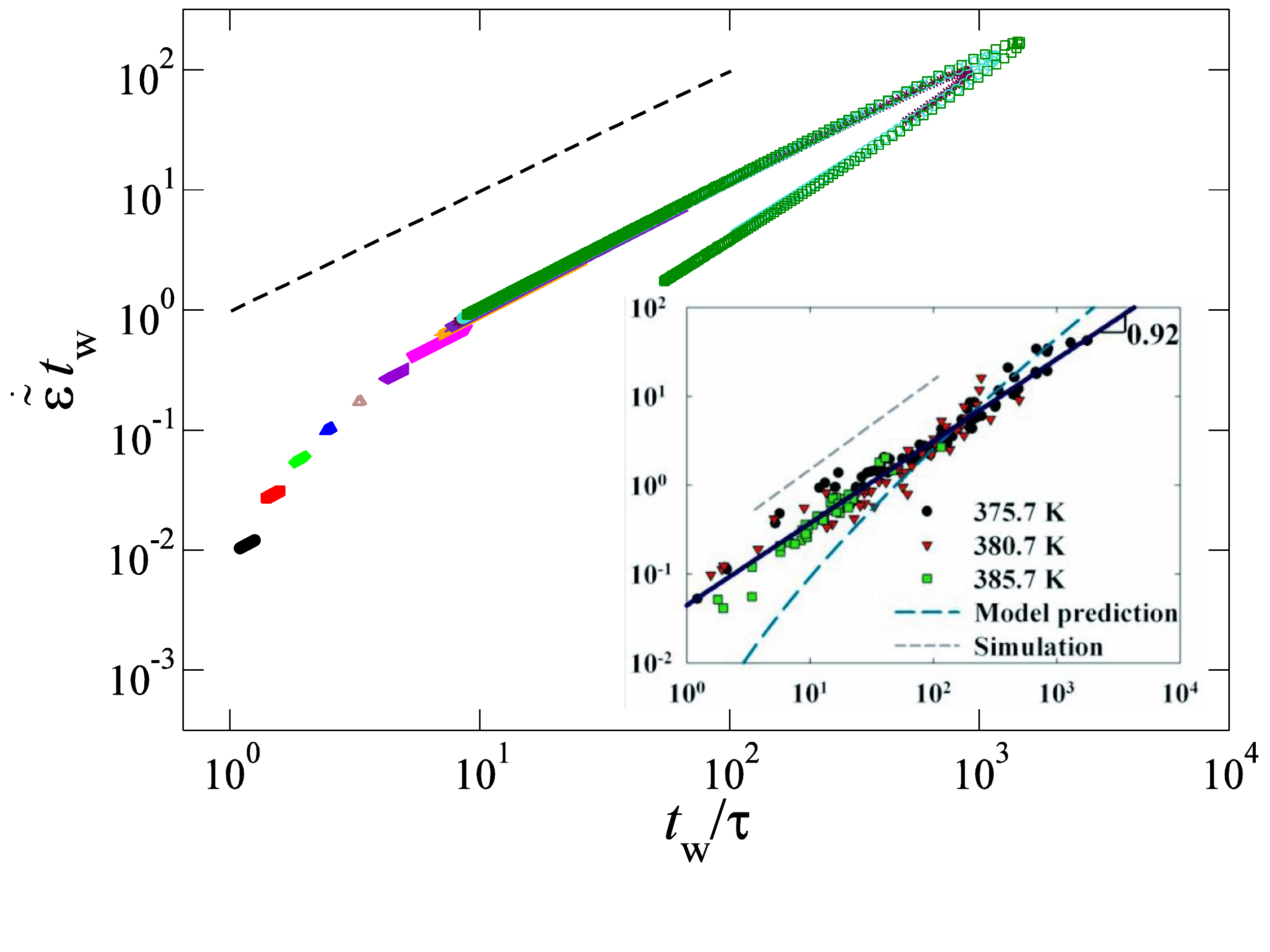}
\caption{(Color online) For the same runs as in Fig.\ref{fig:twoA}, during
loading phase only, log-log plot of reduced strain rate against
reduced relaxation rate. Inset: experimental data collapse
with this plot, slope 0.92. 
(From ~\cite{Ediger}. Reprinted by permission of AAAS.)
Partial collapse occurs here (dashed line is slope 1): while re-entrant regions do not fully superpose, the slopes of the rising and decreasing curves remain comparable.}
\label{fig:twoB}
\end{center}
\end{figure}

We find a good semiquantitative account of the strain curve and $\tau(t)$ data, up to but not beyond the point of unload, by choosing $\tau_0\simeq 6$s. (Unloading is addressed separately below.) Fig.~\ref{fig:one} shows not only the local strain and the segmental relaxation time $\tau(t)$, but also the tensile stresses $T^{p,s}$ carried by polymer and solvent respectively. Key features of the experimental data, reproduced by our minimal model, include: (i) the initial drop in $\tau$ on applying the load; (ii) its subsequent further decline to a state of strong fluidization, with a sharp minimum $\tau_{\rm min}$ near the point of maximum elongation rate; and (iii) its rapid rise from the minimum to a strain-hardened plateau prior to the point of unloading. Not only the initial tenfold drop in $\tau$ on loading but also the subsequent further sharp dip is quantitatively accounted for.
Figure \ref{fig:twoA} shows $\tau$ as a function of the elongational stress, with breakdown of the Eyring-like expectation of a monotonic, single-valued plot.
Figure \ref{fig:twoB} shows (on log-log)
$\dot\varepsilon$ against $1/\tau$; this plot was found to collapse the experimental data in \cite{Ediger} and a similar, if lesser, effect is seen here.  Considering the crudeness of our model (for example, the representation of polymers and solvent by a single mode each), this is remarkable agreement.

If our model is correct, the physics of all these effects is remarkably simple. The (pre-aged) `solvent' glass has a yield stress $\Sigma^s_Y$ (in our model this obeys $\Sigma^s_Y= G^sg(\sqrt{3}\mu)$ with $g(y)\equiv 3y/(y-2)(y+1)$) which is initially exceeded by the applied load. After an initial step-down in $\tau$ caused by step strain on loading, the material yields and progressively fluidizes further; accordingly its strain rate accelerates, giving positive feedback and a collapse in $\tau$. As deformation builds up, however, an ever growing share of the applied stress is instead carried by the stretching polymer chains. This causes the flow rate to drop, so that the solvent, whose stress now obeys $\Sigma^s < \Sigma^s_Y$, starts to solidify. This simple view of strain hardening also directly explains the remarkable behavior of $\tau(t)$.  

Models that factorize aging and rejuvenation effects \cite{EGP} are seriously challenged by the rapid recovery of $\tau$ after the dip. (A multimode spectrum \cite{EGP3} is unlikely to help here.) With simple aging, such factorization predicts $\tau \sim(t+t_w)f(\epsilon)$, so that if the segmental relaxation times falls from its pre-deformation value $t_w$ to a small value $\tau = ft_w = \tau_{min}$ at the dip, a tenfold recovery to $\tau \sim 10 \tau_{min}$ does not occur until $t\sim 10t_w \sim 6\times 10^5$s. This prediction is 100 times too long \cite{Ediger}.

We have also performed numerical calculations in the case of a finite cylinder subject to a necking instability. More details, and an additional figure, are provided in \cite{Methods}. Although our model is not predictive of sample shapes (which depend on the details of the perturbation used to initiate the neck), plots of  $\tau(t)$, and sample radius $\rho(t)$, at three different initial positions along the sample are in qualitative accord with the experiments of  \cite{Ediger}. The explanation given above for the temporal behavior of $\tau(t)$ during elongation of an infinite uniform cylinder remains equally valid for a finite, necked one. 

In order to confirm that our model also behaves reasonably in strain-controlled flows, we have calculated stress responses for startup of steady elongation and compression. These show an overshoot (see additional figure in \cite{Methods}) whose height varies as $\ln(\dot\varepsilon t_w)$, similar to the behaviour found in simple aging fluids \cite{SGR}, and in broad accord with the polymer glass literature.

These successes are  very encouraging. However, the model as formulated so far breaks down badly when the sample is unloaded. Here the experiments show a modest drop in $\tau$ immediately on removing the load, followed by a gradual recovery towards the pre-deformation value. The dotted line in Fig.~\ref{fig:one} shows the prediction based on Eqs.~(\ref{one}--\ref{four}); $\tau$ drops, but then falls much further before recovering. The reason for this behavior within our model is clear. In the strain-hardened regime, the polymers carry a large elastic tensile stress, which exceeds $\Sigma^s_Y$. Upon unloading, this acts backwards on the vitrified solvent, causing it to yield. The resulting $\tau(t)$ resembles a re-run of the initial loading experiment. Another discrepancy is that the value of $G^p \simeq 6$ MPa needed to fit the loading data is approximately ten times larger than the rubbery modulus of the same material above its glass transition (see, e.g., \cite{SHPaper1}). This confirms that the strain-hardened modulus of polymer glasses does not primarily stem from single-chain entropic elasticity  \cite{KS6,Robbins}.

We now identify a physical mechanism that could account for both discrepancies. We invoke the well established phenomenon of large but viscous stresses that arise when chains are strained rapidly relative to their own relaxation time ($\dot\varepsilon\tau^p\gg 1$). Under such conditions, relatively small sections of the polymer quickly stretch close to full extension locally, forming a nearly one-dimensional
multiply folded (`kinked') filament \cite{LarsonKink,Hinch}. Further stretching occurs by migration and annihilation of neighboring kinks of opposite sign. During this process, a large fraction of the stress carried by the polymers is not entropic-elastic, but instead caused by viscous drag against extended subsections of chain. Upon unloading, a large fraction of this inelastic polymer stress disappears on a very rapid timescale \cite{LarsonKink}. 
This mechanism is closely related to the phenomenon of `chain conformation hysteresis' which arises for fully stretched chains, causing a sudden loss of polymer stress on unloading with only modest relaxation of polymer conformations
\cite{SCF}. 

Rather than attempt a full treatment of this rather complex effect (which would certainly require a multimode polymer description \cite{LarsonKink}), we retain our equations
but suppose phenomenologically that the effective polymer modulus drops by a certain factor, $G^p\to \theta G^p$, during unloading of the sample. The solid line in Fig.~\ref{fig:one} shows the result for $\theta = 0.1$. This choice of $\theta$ is consistent with the fitted $G^p$ being ten times larger than the value expected from entropic elasticity alone.
The polymer stress acting backwards on the solvent is now safely below the solvent yield stress; the result is a modest drop and then slow increase in $\tau(t)$, as seen experimentally.

Overall, the success of our simplified model suggests that the striking time dependence of the segmental mobility under elongation, reported in \cite{Ediger}, should be a robustly universal feature of near-$T_g$ polymer glasses. However the quantitative details strongly depend on dimensionless parameters such as $G^s/G^p,\mu$ and $\theta$. We cannot link these directly to microscopic physics, but such parameters can influenced by increasing polymer stiffness, adding small molecules, or introducing short side chains. (All of these should increase $G^s/G^p$, by raising the ratio of solvent-like to polymeric degrees of freedom.) Our model may thus suggest design strategies for manipulating the evolution of $\tau(t)$, tailoring the mechanical responses of polymer glasses to suit particular design needs.

In conclusion, we have presented a simple approach to the modeling of polymer glasses that builds on recent models of rheological aging and rejuvenation in simple glassy fluids. Without attempting to capture every feature of the experiments of \cite{Ediger} (for instance, we do not address the non-exponential form of local relaxations), the minimal combination of a simple-aging fluid with a strain-dependent polymer stress can explain much of what happens when a polymer glass is subjected to elongational load. The unloading behavior is less easily explained, but consistent with a plausible modification of the same model, which crudely allows for the presence of non-elastic polymer stresses when $\dot\varepsilon\tau^p$ is large \cite{LarsonKink}. 

Our work suggests that an accurate representation of aging and rejuvenation physics will form a key part of any more comprehensive theory of polymer glass rheology. 
It encourages the view that a more comprehensive account of polymer glasses might be achieved by judiciously combining existing types of nonlinear rheological theory (describing non-glassy polymers and simple glasses respectively).
Quantitative progress along these lines might enable rapid advances towards the design of superior polymer glass materials.

{\bf Acknowledgements:} MEC is funded by the Royal Society. This work was funded in part by EPSRC EP/E030173. RGL is partially supported from NSF under grant DMR 0906587. Any opinions, findings, and conclusions or recommendations expressed in this material are those of the authors and do not necessarily reflect the views of the National Science Foundation (NSF).

\newpage
\section*{SUPPLEMENTARY MATERIAL}

\setcounter{figure}{0}
\setcounter{equation}{0}
\setcounter{page}{1}

Here we detail the model equations and their transformation via a lubrication approximation into a form suitable for numerical study (by standard methods that we mention only briefly). This is followed by presentation of additional results on necking and on startup flows, and a discussion of the parameter estimates made in relation to the experiments of Lee et al \cite{Ediger}.

\section*{Equations of Motion and Lubrication Approximation}

We adopt Equations (1-3) of the main text plus the continuity  equation for an incompressible fluid ($\nabla.{\bf v} = 0$) and the force balance equation which reads
\begin{equation}
\nabla.[\bS + 2\eta{\bf D}] = 0 
\end{equation}
with the deviatoric stress written as
\begin{equation}
\bS = 
\theta G^p(\bs^p-{\bf I})+G^s(\bs^s-{\bf I})
\end{equation}
Here the phenomenological factor $\theta$ is unity during loading but can drop upon unload as described in the main text. The parameter $\eta$ is a small additional viscosity introduced for purely numerical purposes (to avoid having to deal with inertia), which is always negligible in practice.

To constuct a lubrication approximation (long slender samples) we introduce a coordinate $z$ along the elongation direction; we denote by $x$ a radial coordinate and $A(z,t)$ the cross sectional area. The local fluid velocity along $z$ is denoted $v(z,t)$ so the local elongation rate is $\dot\varepsilon = \partial_z v$. The continuity equation then reads
\begin{equation}
\dot A = - v\partial_zA - A\dot\varepsilon
\end{equation} 
whereas force balance demands
\begin{equation}
\partial_zF = 0
\end{equation} 
where the local tensile force is denoted as
\begin{equation}
F(t) \equiv [A(\Sigma_{zz}-\Sigma_{xx} +3\eta\dot\varepsilon)]
\end{equation} 
(Force balance require this to be independent of $z$.)
In this expression for $F$ we have used the lubrication approximation that velocity gradients depend on $z$ only (not $x$) from which $(\nabla{\bf v})_{zz} = -2 (\nabla{\bf v})_{xx}$ follows by incompressibility \cite{Denn, Pearson,Olagunju}. The same approximation in the Maxwell equations for polymer and solvent (Eqs. 1 and 2 in the main text) gives
\begin{eqnarray}
\dot\sigma^s_{zz} &=& - v\partial_z\sigma^s_{zz} + 2\dot\varepsilon \sigma^s_{zz} - (\sigma^s_{zz}-1)/\tau\\
\dot\sigma^s_{xx} &=& - v\partial_z\sigma^s_{xx} -\dot\varepsilon \sigma^s_{xx} - (\sigma^s_{xx}-1)/\tau\\
\dot\sigma^p_{zz} &=& - v\partial_z\sigma^p_{zz} + 2\dot\varepsilon \sigma^p_{zz} - (\sigma^s_{zz}-1)/\tau_p\\
\dot\sigma^p_{xx} &=& - v\partial_z\sigma^p_{xx} -\dot\varepsilon \sigma^p_{xx} - (\sigma^p_{xx}-1)/\tau_p
\end{eqnarray}
with $\tau_p = \tau/\alpha$ the polymer relaxation time. The equation of motion for $\tau$ (combining Eqs. 3 and 4 of the main text) becomes in the lubrication approximation
\begin{equation}
\dot\tau = -v\partial_z\tau + 1-(\tau-\tau_0)\mu\sqrt{3}|\dot\varepsilon| \label{suptau}
\end{equation}
\section*{Coordinate Transformation}
We now specialize to the case where the applied tensile force $F(t)$ is a piecewise constant function of time. We make a coordinate transformation that removes the exponential increase in sample length (and matching shrinkage of area) corresponding to uniform affine deformation given by the spatially averaged strain ${\overline\varepsilon}(t)$.
This greatly helps the numerical analysis by obviating the need for an adaptive numerical mesh to cope with large sample deformations at late times. The variables $z,v,\dot\epsilon,A,F$ are transformed into a new set $u,w,\dot\zeta,a,f$ as follows:
\begin{eqnarray}
z &=& u \exp[{\overline\varepsilon}(t)]\\
v &=& [\dot{\overline\varepsilon}(t)u + w(u,t)]\exp[{\overline\varepsilon}(t)]\\
\dot\varepsilon &=& \dot{\overline\varepsilon}(t) + \dot\zeta(u,t)\\
A &=& A_0 a(u,t)\exp[-{\overline\varepsilon}(t)]\\
F &=& A_0 f
\end{eqnarray}
where $f$ is (piecewise) constant in time. The remaining variables ($\sigma^s_{xx,zz},\sigma^p_{xx,zz},\tau$) are not transformed, but expressed as functions of ($u,t$) rather than ($z,t$) as previously. Without loss of generality we take the initial length of the sample to be unity; in the absence of necking, the sample domain is $0<u<1$ for all subsequent times. The initial cross sectional area $A_0$ defines the undeformed starting shape of the sample, with $a(u,0) = 1$ for all $u$. 

The transformed equations of motion are
\begin{eqnarray}
f &=& \exp[-\overline\varepsilon(t)]a[\Sigma_{zz}-\Sigma_{xx} + 3\eta (\dot{\overline\varepsilon}+\dot\zeta)]\label{scale1}\\
\dot a &=& - w\partial_u a -\dot\zeta a \label{scale2}\\
\dot\sigma^s_{zz} &=& - w\partial_u\sigma^s_{zz} + 2(\dot{\overline\varepsilon}+\dot\zeta) \sigma^s_{zz} - (\sigma^s_{zz}-1)/\tau\\
\dot\sigma^s_{xx} &=& - w\partial_u\sigma^s_{xx} -(\dot{\overline\varepsilon}+\dot\zeta) \sigma^s_{xx} - (\sigma^s_{xx}-1)/\tau\\
\dot\sigma^p_{zz} &=& - w\partial_u\sigma^p_{zz} + 2(\dot{\overline\varepsilon}+\dot\zeta) \sigma^p_{zz} - (\sigma^s_{zz}-1)/\tau_p\\
\dot\sigma^p_{xx} &=& - w\partial_u\sigma^p_{xx} -(\dot{\overline\varepsilon}+\dot\zeta) \sigma^p_{xx} - (\sigma^p_{xx}-1)/\tau_p\\
\dot\tau &=& - w\partial_u\tau + 1 - (\tau-\tau_0)\mu\sqrt{3}|\dot{\overline\varepsilon}+\dot\zeta|\label{scale7}
\end{eqnarray}
To solve these equations numerically, $a, \sigma^s_{xx,zz},\sigma^p_{xx,zz},\tau$ are updated each time step using Equations \ref{scale2}-\ref{scale7}. Updated values are fed into Equation \ref{scale1}, whose integral over the entire sample determines $\dot{\overline\varepsilon}$; this value is then substituted back to determine $\dot\zeta(u,t)$ everywhere. 

If one assumes the sample does not neck (so that the deformation remains affine) these equations reduce to ordinary differential equations: quantities depend on $t$ but not $u$, and by construction, $\dot\zeta = 0$. 
Numerical solution proceeds by an explicit Euler algorithm \cite{recipes}, with careful attention paid to timestep convergence.
In our numerics we chose $\alpha = 10^{-12}$, but the results are robust to variations at small $\alpha$.
We choose $\eta = 0.05$, which, as previously indicated is effectively zero and the same applies. 

\begin{figure}
\begin{center}
\includegraphics[width=85mm]{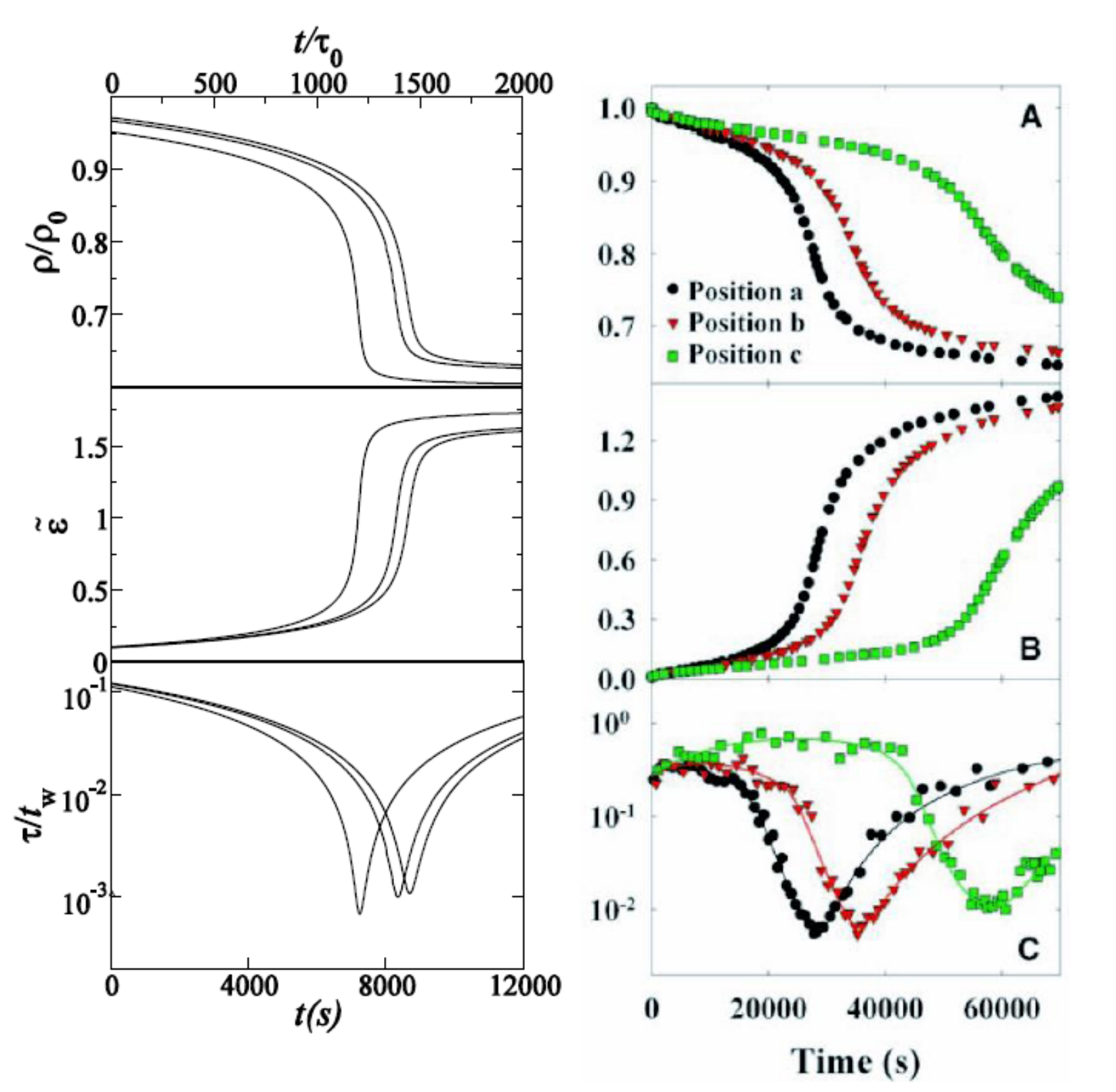}
\caption{ (Color online) Left Panels: the reduced sample radius
$\rho/\rho_0$ (with $\rho_0$ the initial value), the local strain
measure $\tilde{\varepsilon}=\exp\varepsilon-1$ \cite{Ediger}, and the relaxation time $\tau(t)$ during sample
loading for three different positions in a finite cylinder subject to
necking instability. Parameters as in Fig.1 of main text, now for a cylinder that
initially occupies the space $z=0$ to $z=1$, spatially resolved using
$400$ numerical grid points. The three positions chosen are initially
at $z=0.5,0.9,0.95$. Right Panels: Similar experimental data; note the discrepancy in horizontal scales (no such discrepancy arose for Figs.1-3 of the main text). (From ~\cite{Ediger}. Reprinted by permission of AAAS.)}
\label{fig:three}
\end{center}
\end{figure}

\begin{figure}
\begin{center}
\includegraphics[width=70mm]{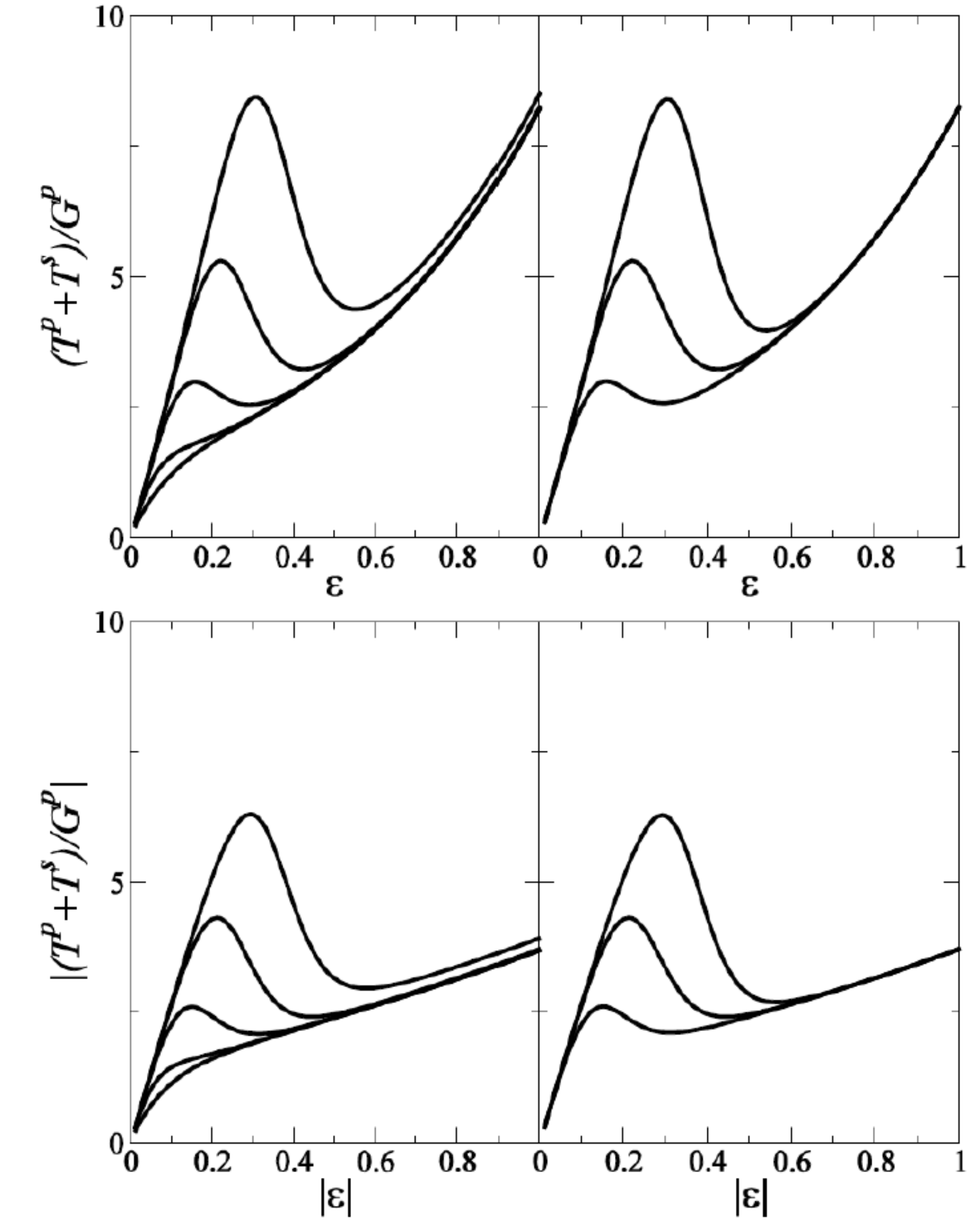}
\caption{
Upper Panels: Dimensionless tensile stress $(T^p+T^s)/G^p$ versus Hencky strain $\dot\varepsilon t$ in startup of steady elongation at strain rate $\dot\varepsilon$. Left $t_w/\tau_0 = 10^4$ with (bottom to top) $\dot\varepsilon\tau_0 = 10^{-6}, 10^{-5}, 10^{-4}, 10^{-3}, 10^{-2}$. Right, $\dot\varepsilon\tau_0 = 10^{-3}$ with (bottom to top) $t_w/\tau_0 = 10^3, 10^4, 10^5$. Lower panels: The same, for startup of compressional flow ($\dot\varepsilon<0$). The remaining model parameters are as in Fig.1 of main text (except that the (unimportant) numerical viscosity $\eta$ is set equal to zero).  \label{fig:S1}}
\end{center}
\end{figure}

\section*{Necking}

In general, however, one expects non-affine deformation of the sample due to necking. The partial differential equations (\ref{scale1}--\ref{scale7}) must then be solved in full. For this we split the operator, using explicit Euler for the local terms and first order upwinding for the convective ones \cite{recipes}.
In addition, it is now necessary to specify boundary conditions, and also to introduce some perturbation to the affine state so as to trigger the necking instability. In practice, necking occurs not in a random position but near the middle of the sample -- presumably because clamping or other boundary conditions at the extremities break the translational symmetry. Retaining periodic boundary conditions for numerical simplicity, we add an early time perturbation of the form 
\begin{equation}
\frac{\dot a}{\dot{\overline\varepsilon}+\dot\zeta} = [\exp(-u^2{\exp(2\overline\varepsilon)}/A_0)
+\exp(-(1-u)^2{\exp(2\overline\varepsilon)}/A_0)]
\label{neckpert}
\end{equation}
whose effect is to create a sample comprising a long cylindrical section that flares out at both ends on a length scale $\sim \sqrt{A_0}$ comparable to the sample width. (This is a plausible representation of the early-time shape of a sample that is not yet necked but whose ends, for whatever reason, have an area different from the affine value.) This perturbation is applied as an additional term in Eq. \ref{scale2}, but only at early times; it is switched off once the mean strains $\overline\varepsilon$ exceeds a chosen threshold, $\varepsilon^*$. 

Fig. \ref{fig:three} shows results for the sample radius and segmental relaxation time at three different positions in a necking cylinder, computed within the lubrication theory described above.

\section*{Strain-Controlled Flows}

Figure \ref{fig:S1} shows our predictions (for a non-necking infinite cylinder) for the stress overshoot in polymer glasses in startup flow.
We are not aware of startup data on the corresponding experimental system as studied by \cite{Ediger}.
Nonetheless these curves qualitatively resemble literature data on startup in polymer glasses (see, e.g.,  Fig. 1 of \cite{EGP}). Our model predicts a logarithmic dependence of overshoot amplitude on both strain rate and sample age with roughly a two-fold overshoot increase for a ten-fold increase in age or strain rate. (This ratio may depend on model details.) The overshoot is controlled mainly by the underlying fluidity model rather than polymeric effects.

\section*{Parameter Estimation}

As mentioned in the main text we set $\alpha = 10^{-12}$, which is effectively zero, and set the dimensionless aging parameter $d\tau/dt$ (in a system at rest) to be unity for simplicity. We have already chosen our length unit as the initial sample length. The parameters remaining in the model are the material parameters $G^p,G^s,\tau_0,\mu$; the tensile force $F$; the waiting time $t_w$; the unloading factor $\theta$; the perturbation threshold $\varepsilon^*$ introduced above; the initial cross sectional area of the sample $A_0$; and the additional viscosity $\eta$.

As stated previously, the viscosity $\eta$ is a numerical device that allows force balance to be maintained from one time step to the next by slight adjustments of the strain rate. (An alternative would be to include fluid inertia; another would be to have an extra iteration loop to maintain precise force balance at every timestep.) Any small value for $\eta$ will achieve this without corrupting the physics of the model; we choose $\eta = 0.05 G^p\tau_0$. Since $G^s>G^p$ and $\tau>\tau_0$, this always remains negligible compared to the solvent viscosity. 

The initial cross section $A_0$ can in principle be matched to the samples reported by Lee et al \cite{Ediger}, but in a homogeneous deformation none of the quantities we report depend on sample shape. Hence $A_0$ is relevant only when necking is accounted for; when a definite value is needed (as in Eq.\ref{neckpert} above) we choose $A_0=0.1$. (Recall that the unit of length is the sample length). Similar remarks apply to $\varepsilon^*$; for definiteness we choose $\varepsilon^*= 0.04$.

The applied tensile force $F= A_0f$ can be estimated by hypothesizing that, in the plateau that is reached after strain hardening, this force is carried predominantly by the polymers. Thus $F =A(t)G^p(\sigma^p_{zz}-\sigma^p_{xx}) \simeq A(t)G^p\exp (2\varepsilon)$; it follows that $f =G^p\exp(\varepsilon)$.
Lee et al \cite{Ediger} report deformations in terms of a local strain variable $(L(t)-L_0)/L_0$ which equates to $\exp[\varepsilon(t)]-1$ in our notation. Mildly extrapolating the data of their Fig.1 to estimate the plateau height, we conclude that $\exp[\varepsilon(t)]\simeq 2.7$; it follows that $f=2.7 G^p$.

Note that $f \equiv F/A_0$ is in fact the `engineering stress' of \cite{Ediger}, which is 16 MPa in the loading experiment reported in their Fig.1. The resulting $G^p$ of $\sim 6$ MPa is about ten times larger than the rubbery modulus of the same material \cite{SHPaper1}. This is consistent with the view of \cite{KS6} that the modulus responsible for strain hardening, though attributable to polymers, is not simply entropic in origin as in rubbers (see main text). 

We turn next to the parameter $\mu \equiv \lambda/(\sqrt{3}|\dot\varepsilon|)$ which controls the strain-induced relaxation of $\tau$. Within our model, this is fixed by observing that in steady state at low elongation rates ($|\dot\varepsilon|\tau_0\ll1$)
the solution to Eq.\ref{suptau} approaches
\begin{equation}
\tau_{ss}^{-1} = |\dot\varepsilon| \sqrt{3} \mu
\end{equation}
Figure 3 of Lee et al \cite{Ediger} plots $\dot\varepsilon\tau(0)$ against $\tau(0)/\tau_{ss}$ (with $\tau(0)\gg \tau_0$ the pre-deformation value of the relaxation time); the slope on this log-log plot is close to unity. Estimating the near-constant ratio $\dot\varepsilon/\tau_{ss}^{-1}$ from the left edge of the experimental plot gives $\dot\varepsilon = 0.045/\tau_{ss}$ from which we deduce that $\mu \sim 12.5$.

Armed with this value, the solvent modulus $G^s$ is now estimated from the virtually instantaneous drop from the undeformed value of $\tau = \tau(0)$ to a smaller value $\tau^+ \simeq 0.1\tau(0)$, reported immediately after imposition of the load. The resulting strain $\varepsilon^+$ is not directly reported in \cite{Ediger} but by integrating our equations of motion is found to obey
\begin{equation}
f\exp[\varepsilon^+] = (G^p+G^s) (\exp[2\varepsilon^+]-\exp[-\varepsilon^+])
\end{equation}
so that for small strain and large $G^s/G^p$ we have $G^s = f/(3\varepsilon^+)$. 
Integrating the $\tau$ equation through the same step strain, we find the post-load value
$\tau^+ = \tau(0)\exp[-\sqrt{3}\mu\varepsilon^+] =
\tau(0) \exp[-f\mu/\sqrt{3}G^s]$.
Figure 1 in \cite{Ediger} shows $\tau^+/\tau(0)\simeq 0.1$ from which it follows that $G^s/G^p = 8.5$ (using the previous result that $f = 2.7 G^p$). 

The resulting $G^s\sim 50$ MPa is $30-50$ times smaller than the Young's modulus of fully vitrified PMMA \cite{GlassMod}, which suggests that in the temperature range of interest (near but below the glass transition) only a small proportion of the monomeric degrees of freedom are directly involved in the process of solvent arrest. 
 
The next parameters to be estimated are $t_w$, the effective age of the sample on first loading, and $\tau_0$ which is the relaxation time of a fully devitrified sample. Within our model, the initial relaxation time $\tau(0)$ is simply equal to $t_w$.
The remaining parameter $\tau_0$ is then needed to connect the time axis in our numerical calculations to that in the experiments of \cite{Ediger}. In essence, we use this as an implicit fit parameter to get the best agreement between our $\tau(t)$ curves and those reported experimentally. (In numerical practice, we use $\tau_0$ as the unknown but fixed unit of time in which case $t_w/\tau_0$, the numerical sample age, is the explicit fit parameter.)  To extract a value of $\tau_0$  from our fits we observe that, when the loaded sample in Figure 1 of \cite{Ediger} achieves its minimum relaxation time (point (e) on panel C of that figure), the decay curve from which this time is measured (panel B) is close to mono-exponential with a relaxation time of 28s. The relaxation time $\tau(0)$ for the undeformed material (for which the decay is far from exponential but fitted carefully by the authors of \cite{Ediger}) is (from their Figure 1 A) $10^{3.3}$ times larger, i.e., $6\times 10^4$ s. Our best fit waiting time obeys $t_w/\tau_0 = 10^4$; since $\tau(0) = t_w$, we then have $\tau_0=6$ s.

As noted in the main text, the elongational dynamic yield stress of the solvent fluid can be found by calculating the steady state stresses at fixed $\dot\varepsilon$ and studying the limiting tensile stress $T$ as $\dot\varepsilon\to 0$. The result is $\Sigma^s_Y = \sqrt{3} G^s/\mu \simeq 7$ MPa. This is somewhat at odds with Figure 2 of \cite{Ediger}, which suggests an elastic response up to stresses of about twice this value. However, the true yield stress could lie below the value extracted from the experimental measurements if the latter do not detect extremely slow creeping motion that may arise for stresses close to but above $\Sigma^s_Y$. 

The above considerations fix all the parameters of the model except for the unloading factor $\theta$ which was addressed already in the main text.


\begin{thebibliography}{}



\bibitem{McLeishReview} T. C. B. McLeish, 
Adv. in Phys. {\bf 51} 1379-1527 (2002).

\bibitem{LarsonBook} R. G. Larson, {\em Constitutive Equations for Polymer Melts and Solutions} (Butterworth-Heinemann, Boston, 1988).

\bibitem{Ediger}
H.-N. Lee, K. Paeng, S. F. Swallen, M. D. Ediger, 
Science {\bf 323}, 231-234 (2009).


\bibitem{Struik} L. C. E. Struik, {\em Physical Aging in Amorphous Polymers and Other Materials} (Elsevier, New York, 1978).

\bibitem{Molecular} P. G. Debenedetti, F. H. Stillinger, 
Nature {\bf 410}, 259-267 (2001).

\bibitem{Metallic} D. C. Hofmann et al, 
Nature {\bf 451}, 1085-1089 (2008). 

\bibitem{Colloidal} P. Schall, D. A. Weitz, F. Spaepen, 
Science {\bf 318}, 1895-1899 (2007).

\bibitem{BraderPRL2} J. M. Brader, M. E. Cates, M. Fuchs, 
Phys. Rev. Lett. {\bf 101}, 138301 (2008).

\bibitem{BraderPNAS} J. M. Brader, T. Voigtmann, M. Fuchs, R. Larson, M. E. Cates, 
Proc. Nat. Acad. Sci. USA {\bf 106}, 15186-15191 (2009).

\bibitem{SGR} S. M. Fielding, P. Sollich, M. E. Cates, 
J. Rheol. {\bf 44} 323-369 (2000).


\bibitem{KS1} K. Chen, K. S. Schweizer, 
EPL {\bf 79}, 26006 (2007).

\bibitem{STZ} M. L. Falk, J. S. Langer, 
Annu. Rev. Cond. Mat. Phys. {\bf 2}, 353-373 (2010).

\bibitem{Fluidity} C. Derec, A. Ajdari, F. Lequeux, 
Eur. Phys. J. E {\bf 4}, 355-361 (2001).

\bibitem{Fluidity2} P. Coussot, Q. D. Nguyen, H. T. Huynh, D. Bonn, 
Phys. Rev. Lett. {\bf 88} 175501 (2002).

\bibitem{Ediger2} H.-N. Lee and M. D. Ediger, 
Macromolecules {\bf 43} 5863-5873 (2010).

\bibitem{Aging} B. Rinn, P. Maass, J.-P. Bouchaud, 
Phys. Rev. Lett. {\bf 84}, 5403-5406 (2000).

\bibitem{Simulations1} R. A. Riggleman, H. N. Lee, M. D. Ediger, J. J. de Pablo, 
Phys. Rev. Lett. {\bf 99}, 215501 (2007).

\bibitem{Simulations2} H. N. Lee, R. A. Riggleman, J. J. de Pablo, M. D. Ediger, 
Macromolecules {\bf 42}, 4328-4336 (2009).

\bibitem{Simulations3} M. Warren, J. Rottler, 
Phys. Rev. E {\bf 76}, 031802 (2007).

\bibitem{Simulations4} M. Warren, J. Rottler, 
Phys. Rev. Lett. {\bf 104}, 205501 (2010).

\bibitem{Eyring} H. Eyring, 
J. Chem. Phys {\bf 4}, 283-291 (1936).

\bibitem{KS2} K. Chen, K. S. Schweizer, 
Phys. Rev. Lett. {\bf 102}, 038301 (2009);

\bibitem{KS3} 
K. Chen, K. S. Schweizer, 
Phys. Rev. E {\bf 82}, 041804 (2010).

\bibitem{KS6} K. Chen, E. J. Saltzman, K. S. Schweizer, 
J. Phys. Cond. Mat. {\bf 21} 503101 (2009).

\bibitem{EGP2} R. N. Haward and G. Thackray, 
Proc. Roy. Soc. Lond. Ser. A {\bf 302} 453-372 (1968).

\bibitem{EGP} E. T. J. Klompen, T. A. P. Engels, L. E. Govaert and H. E. H. Meijer, 
Macromolecules {\bf 38} 6997-7008 (2005).

\bibitem{hoy} R. S. Hoy and C. S. O'Hern, 
Phys. Rev. E {\bf 82}, 041803 (2010).

\bibitem{Robbins} R. S. Hoy, M. O. Robbins, 
Phys. Rev. Lett. {\bf 99} 117801 (2007).

\bibitem{Methods} The lubrication theory, numerical methods, parameter estimations, and some additional results, are described online at XXXX XXXX.

\bibitem{EGP3} L. C. A. van Breemen, E. T. J. Klompen, L. E. Govaert and H. E. H. Meijer, 
J. Mech. Phys. Solids, in press (2011).

\bibitem{SHPaper1} R. N. Haward, 
Macromolecules {\bf 26}, 5860-5869 (1993).

\bibitem{LarsonKink} R. G. Larson, 
Rheol. Acta {\bf 29} 371-384 (1990).

\bibitem{Hinch} E. J. Hinch, 
J. Non-Newtonian Fluid Mech. {\bf 54}, 209-230 (1994).

\bibitem{SCF} J. P. Rothstein, G. H. McKinley, 
J. Non-Newtonian Fluid Mech. {\bf 108}, 275-290 (2002).

\bibitem{Denn} M. M. Denn, 
Ann. Rev. Fluid. Mech. {\bf 12}, 365-387 (1980).

\bibitem{Pearson} M. A. Matovich, J. R. A. Pearson, 
I and EC Fundamentals {\bf 8}, 512 (1969).

\bibitem{Olagunju} D. O. Olagunju, 
J. Non-Newtonian Fluid Mech. {\bf 87}, 27-46 (1999).

\bibitem{recipes} W. H. Press {\em et al.}, {\em Numerical Recipes in C} (Cambridge University Press, Cambridge 1988).

\bibitem{GlassMod} J. E. Mark, Ed., {\em Physical Properties of Polymers Handbook} 2nd Edition (Springer, NY, 2007).

\end{thebibliography}
\end{document}